\documentclass[aps,pra,twocolumn,reprint,amsmath,amssymb,superscriptaddress,floatfix,footinbib,longbibliography]{revtex4-1}

\usepackage{float}

\usepackage{epsfig}
\usepackage{graphicx}
\usepackage{epstopdf}

\usepackage[T1]{fontenc}
\usepackage[applemac]{inputenc}
\usepackage{lmodern}
\usepackage[english]{babel}

\usepackage{ae}
\usepackage{units}

\usepackage{amsmath,amssymb,natbib,bm}
\usepackage{psfrag}
\usepackage{subfigure}
\usepackage{amsthm}

\usepackage{color}
\usepackage{url}

\usepackage[colorlinks]{hyperref}
\hypersetup{%
        plainpages=true,
        breaklinks=true,
        hypertexnames=false,
        pageanchor=true,
        colorlinks=true,
        linkcolor={blue},
        citecolor={magenta},
        urlcolor={blue},
        anchorcolor={black}
      }
      
\usepackage{mleftright} %to get rid of the extra space around parentheses with \left and \right: write \mleft, \mright instead

\newcommand{\figref}[1]{\mbox{Fig.~\ref{#1}}}
\newcommand{\tabref}[1]{\mbox{Table~\ref{#1}}}
\newcommand{\secref}[1]{\mbox{Sec.~\ref{#1}}}

\newcommand{\appref}[1]{\mbox{Appendix~\ref{#1}}}
\renewcommand{\eqref}[1]{\mbox{Eq.~(\ref{#1})}}

\newcommand{\ket}[1]{|#1\rangle}

\newcommand{\ketbra}[2]{\mleft| #1 \rangle \langle #2 \mright|}
\newcommand{\brakket}[3]{\langle #1 | #2 | #3 \rangle}
\newcommand{\expec}[1]{\mleft\langle #1 \mright\rangle}

\newcommand{\comm}[2]{\mleft[ #1, #2 \mright]}
\newcommand{\lind}[1]{\mathcal{D}\mleft[#1\mright]}

\newcommand{\sz}{\hat \sigma_z}
\newcommand{\sx}{\hat \sigma_x}

\newcommand{\nn}{\nonumber}

\newcommand{\be}{\begin{equation}}
\newcommand{\ee}{\end{equation}}
\newcommand{\bea}{\begin{eqnarray}}
\newcommand{\eea}{\end{eqnarray}}

\begin{document}

\title{Simple preparation of Bell and GHZ states using ultrastrong-coupling circuit QED}

\author{Vincenzo Macr\`{i}}
\affiliation{Theoretical Quantum Physics Laboratory, RIKEN Cluster for Pioneering Research, Wako-shi, Saitama 351-0198, Japan}

\author{Franco Nori}
\affiliation{Theoretical Quantum Physics Laboratory, RIKEN Cluster for Pioneering Research, Wako-shi, Saitama 351-0198, Japan}
\affiliation{Physics Department, The University of Michigan, Ann Arbor, Michigan 48109-1040, USA}

\author{Anton Frisk Kockum}
\email[e-mail:]{anton.frisk.kockum@gmail.com}
\affiliation{Theoretical Quantum Physics Laboratory, RIKEN Cluster for Pioneering Research, Wako-shi, Saitama 351-0198, Japan}
\affiliation{Wallenberg Centre for Quantum Technology, Department of Microtechnology and Nanoscience, Chalmers University of Technology, 412 96 Gothenburg, Sweden}

\date{\today}

\begin{abstract}

The ability to entangle quantum systems is crucial for many applications in quantum technology, including quantum communication and quantum computing. Here, we propose a new, simple, and versatile setup for deterministically creating Bell and Greenberger-Horne-Zeilinger (GHZ) states between photons of different frequencies in a two-step protocol. The setup consists of a quantum bit (qubit) coupled ultrastrongly to three photonic resonator modes. The only operations needed in our protocol are to put the qubit in a superposition state, and then tune its frequency in and out of resonance with sums of the resonator-mode frequencies. By choosing which frequency we tune the qubit to, we select which entangled state we create. We show that our protocol can be implemented with high fidelity using feasible experimental parameters in state-of-the-art circuit quantum electrodynamics. One possible application of our setup is as a node distributing entanglement in a quantum network.

\end{abstract}

\maketitle

%%%%%%%%%%%%%%%%%%%%%%%%%%%%%%%%%%%%%%%%%%%%%%%%%

\section{Introduction}

Quantum entanglement~\cite{Horodecki2009} plays a key role in quantum communication~\cite{Gisin2007}, quantum computing~\cite{Nielsen2000, Ladd2010, Buluta2011}, and other quantum information processing~\cite{Wendin2017, Gu2017, Flamini2018}. To give just a few examples, quantum teleportation~\cite{Bennett1993}, quantum key distribution~\cite{Ekert1991}, quantum secret sharing~\cite{Hillery1999}, quantum secure direct communication~\cite{Long2002, Deng2003}, and quantum repeaters~\cite{Briegel1998, Li2015}, are some of the quantum communication protocols that require entangling quantum systems.

The simplest examples of entangled states are known as Bell states. They are four maximally entangled states involving two quantum bits (qubits, two-level systems with ground state $\ket{0}$ and excited state $\ket{1}$):
\bea
\ket{\Phi_{\pm}} &=& \frac{1}{\sqrt{2}} \mleft( \ket{00} \pm \ket{11} \mright), \label{eq:BellPhi}\\
\ket{\Psi_{\pm}} &=& \frac{1}{\sqrt{2}} \mleft( \ket{01} \pm \ket{10} \mright).
\eea
The Bell states are of fundamental importance in both quantum cryptography and quantum teleportation. With $N \geq 3$ qubits, maximally entangled states such as the Greenberger-Horne-Zeilinger (GHZ) states~\cite{Greenberger1989, Bouwmeester1999},
\be
\ket{\Psi_{\rm GHZ}} = \frac{1}{\sqrt{2}} \mleft( \ket{0}^{\otimes N} + \ket{1}^{\otimes N} \mright),
\ee
and the W states~\cite{Dur2000}, 
\be
\ket{\Psi_{\rm W}} = \frac{1}{\sqrt{N}} \mleft( \ket{100 \ldots 0} + \ket{010 \ldots 0} + \ldots + \ket{000 \ldots 1} \mright),
\ee
are not only of intrinsic interest but also of great practical importance. New systems and methods for preparing and measuring such entangled states has therefore been sought intensively for a long time~\cite{Aspect1982, Weihs1998, Molmer1999, Pan2000, Zhao2003, Wei2006, Steffen2006, Coelho2009, Bishop2009, Zhang2009, DiCarlo2010, Wang2010, Hasegawa2010, Pan2012, Riste2013a, Su2014, Johansson2014, Zang2015, Yang2016, Kang2016, Tashima2016, Zhang2016, Ritboon2017, Erhard2017a, Qin2018, Cruz2018}, and remains a very active field of research. In recent years, entanglement of ten or more qubits has been demonstrated in various experimental setups~\cite{Monz2011, Wang2016, Hou2016, Song2017, Wang2018}.

In this article, we propose a simple method for the deterministic preparation of Bell and GHZ states using ultrastrong coupling (USC)~\cite{Kockum2018a, Forn-Diaz2018} between light and matter. In this regime of light-matter interaction, the coupling strength $g$ becomes comparable to the bare transition frequencies $\omega$ in the system. In the past decade, USC has been realized in several experimental systems~\cite{Kockum2018a}, including intersubband polaritons~\cite{Anappara2009, Gunter2009, Geiser2012, Askenazi2017} and Landau polaritons~\cite{Muravev2011, Scalari2012, Bayer2017} in quantum wells, superconducting circuits~\cite{Niemczyk2010, Forn-Diaz2010, Chen2017, Yoshihara2017, Forn-Diaz2017}, organic cavities in photonic cavities~\cite{Schwartz2011, Gambino2014, Barachati2018, Genco2018}, and optomechanical systems~\cite{George2016a, Benz2016}. Out of these systems, we believe our proposal is most suited for superconducting circuits, i.e., the circuit version of cavity quantum electrodynamics (QED)~\cite{Haroche2013} known as circuit QED~\cite{You2011, Gu2017}. The reason for this is that the circuit-QED experiments are the only ones that have demonstrated USC with \textit{single} (although artificial) atoms.

Although USC leads to much interesting physics~\cite{DeLiberato2007, DeLiberato2009, Ashhab2010, Casanova2010, Ai2010, Cao2010, Cao2011, Stassi2013, Ridolfo2013, Garziano2013, Garziano2014, Huang2014, Sanchez-Burillo2014, DeLiberato2014, Zhao2015, Lolli2015, Cirio2016, DiStefano2017, Macri2018}, in this article, we are only using the fact that it enables higher-order processes that do not conserve the number of excitations in the system~\cite{Zhu2013, Ma2015, Garziano2015, Garziano2016, Kockum2017a, Stassi2017, Kockum2017}. These processes include multiphoton Rabi oscillations~\cite{Garziano2015}, a single photon exciting two spatially separated qubits~\cite{Garziano2016}, and analogues of almost all nonlinear-optics phenomena~\cite{Kockum2017a}, including various frequency-conversion schemes~\cite{Kockum2017}.

Our proposal for generating Bell and GHZ states builds on ideas from Refs.~\cite{Kockum2017, Stassi2017}. In Ref.~\cite{Kockum2017}, we showed how to realize frequency conversion of photons in two resonators (or resonator modes) ultrastrongly coupled to a single qubits, and that these processes can be well controlled by tuning the qubit frequency in and out of resonance conditions for these processes. The ability to tune the qubit frequency in this way is available in many circuit-QED experiments. In the present work, we consider setups with the qubit coupled ultrastrongly to two or three resonator modes. The key difference to Ref.~\cite{Kockum2017}, which allows us to create various entangled states between photons in the different resonator modes, is that we first prepare the qubit in a superposition, and then use frequency-conversion processes to transfer this superposition to the photons. 

For example, to prepare a Bell state with photons in the first two resonator modes, we first prepare the qubit in a superposition state with equal amplitudes for being in the ground state and for being in the excited state. We then tune the qubit frequency to equal the sum of the transition frequencies for the two resonator modes. This resonance condition enables a higher-order process that transfers the qubit excitation to photons in the resonator modes and back in a Rabi oscillation. By detuning the qubit just when the excitation is fully in the resonator modes, we end up with a Bell state of the type shown in \eqref{eq:BellPhi}.

Our proposed setup is a \textit{simple} and \textit{versatile} entanglement generator. The only operations required are to prepare one qubit in a superposition state and then tuning it in and out of resonance. The same setup can both generate Bell states for any pair of resonator modes and GHZ states for all the modes. The only part of the protocol that needs to be adjusted, to choose which entangled state to generate, is which value the qubit frequency is tuned to. This could be useful in quantum information processing, e.g., for distributing entanglement at a node in a quantum network~\cite{Kimble2008, Wehner2018}. Our setup is also versatile in the sense that we can generate entanglement between photons of several different colors (the frequencies are given by the resonator-mode frequencies); it could be said that we create ``rainbow'' entangled states, similar to Ref.~\cite{Coelho2009}.

This article is organized as follows. In \secref{sec:Model}, we describe our system in detail. We plot the energy levels of the system Hamiltonian to illustrate how tuning the qubit frequency enables the different entangling processes we want, and we give analytical expressions for the effective interaction strength for these processes, which sets the time needed to create the entangled state. We also describe how we model losses in the system. In \secref{sec:Results}, we explain the details of our protocol for entanglement generation and present results of full numerical simulations of these protocols, using experimentally feasible parameters and including losses of varying degree. We then conclude and give an outlook for future work and applications in \secref{sec:ConclusionOutlook}. The analytical calculations for the effective interaction strengths are presented in detail in Appendices~\ref{app:CalculationsBell} and \ref{app:CalculationsGHZ}.

%%%%%%%%%%%%%%%%%%%%%%%%%%%%%%%%%%%%%%%%%%%%%%%%%

\section{Description of the model}
\label{sec:Model}

%%%%%%%%%%%%%%%%%%%%%%%%

\subsection{Hamiltonian}

We consider a quantum system consisting of three non-degenerate resonator modes (labelled $a$, $b$, $c$) coupled ultrastrongly to a two-level system (a qubit, labelled $q$), possibly with symmetry-broken potentials. The Hamiltonian describing this system is the generalized quantum Rabi Hamiltonian ($\hbar =1$ throughout this article)
\bea
\hat H &=& \omega_a \hat a^\dag \hat a + \omega_b \hat b^\dag \hat b + \omega_c \hat c^\dag \hat c + \frac{\omega_q}{2} \sz \nn\\
&&+ \mleft[ g_a \mleft( \hat a^\dag + \hat a \mright) + g_b \mleft( \hat b^\dag + \hat b \mright) + g_c \mleft( \hat c^\dag + \hat c \mright) \mright] \nn\\
&&\times \mleft( \sx \cos \theta + \sz \sin \theta \mright) ,
\label{eq:Hamiltonian}
\eea
where $\omega_j$ is the transition frequency of resonator mode $j$, $\omega_q$ is the qubit frequency, and $g_j$ is the strength of the coupling between resonator $j$ and the qubit. The operators $\hat a$, $\hat b$, and $\hat c$ ($\hat a^\dag$, $\hat b^\dag$, and $\hat c^\dag$) are the annihilation (creation) operators of the resonator modes $a$, $b$, and $c$, respectively. The qubit degrees of freedom are described by the Pauli matrices $\sz$ and $\sx$. The angle $\theta$ parameterizes the amount of longitudinal and transversal coupling between the qubit and the resonators.

This mix of longitudinal and transversal coupling can be realized in circuit-QED experiments with flux qubits~\cite{You2007, Deppe2008, Forn-Diaz2010, Niemczyk2010, Baust2016, Yoshihara2017}. Note that the presence of the longitudinal coupling term in \eqref{eq:Hamiltonian} is necessary to generate photonic Bell states in our scheme, since that requires converting one qubit excitation into two photons, which neither conserves the number of excitations in the system nor their parity. However, to generate photonic GHZ states, the transversal coupling, which conserves parity, is sufficient, since in this case one qubit excitation is converted into three photons. Thus, if one only wishes to generate GHZ states, the standard quantum Rabi Hamiltonian ($\theta = 0$) for multiple resonator modes can be used.

%%%%%%%%%%%%%%%%%%%%%%%%

\subsection{Master equation and numerical methods} 

To include the effect of decoherence in our system, we use a master equation on the Lindblad form in our numerical simulations. Following Refs.~\cite{Breuer2002, Beaudoin2011, Ridolfo2012}, we express the system-bath interaction Hamiltonian in the basis formed by the energy eigenstates of $\hat H$ from \eqref{eq:Hamiltonian}. By applying the standard Markov approximation and tracing out the reservoir degrees of freedom, we arrive at the master equation for the density-matrix operator $\hat \rho (t)$,
\bea
\dot{\hat\rho}(t) &=& -i \comm{\hat H}{\hat \rho (t)} + \kappa_a \lind{\hat X_a^-} \hat \rho (t) + \kappa_b \lind{\hat X_b^-} \hat \rho (t) \nn\\
&&+ \kappa_c \lind{\hat X_c^-} \hat \rho (t) + \gamma \lind{\hat S^-} \hat \rho (t) ,
\label{eq:ME}
\eea
where the constants $\kappa_{\rm a}, \kappa_{\rm b}, \kappa_{\rm c}$ and $\gamma$ correspond to the damping rates of the resonator modes and the qubit, respectively. The superoperator ${\cal D}$ is defined as
\be
\lind{\hat O} \hat \rho = \frac{1}{2} \mleft( 2 \hat O \hat \rho \hat O^\dag - \hat O^\dag \hat O \hat \rho - \hat \rho \hat O^\dag \hat O \mright),
\ee
and the dressed lowering operators $\hat O = \hat X_a^-, \hat X_b^-, \hat X_c^-, \hat S^-$ are defined in terms of their bare counterparts $\hat o = \hat a, \hat b, \hat c, \hat \sigma_-$ as~\cite{Ridolfo2012}
\be
\hat O = \sum_{E_n > E_m} \brakket{\Psi_m}{(\hat o + \hat o^\dag)}{\Psi_n} \ketbra{\Psi_m}{\Psi_n} ,
\label{eq:DressedOperators}
\ee
where $\ket{\Psi_n}$ ($n \in \mathbb{N}$) are the eigenvectors of $\hat H$ and $E_n$ the corresponding eigenvalues. Note that in the USC regime, using the bare operators directly in the master equation leads to unphysical effects, such as eternal production of photons from the ground state of the system~\cite{Beaudoin2011, Ridolfo2012}.

In writing the master equation, we have assumed that the environment that the system interacts with is at zero temperature, $T=0$. If needed to better model experiments, the master equation can be extended to account for non-zero temperatures~\cite{Settineri2018}.

The spectrum and the eigenstates of $\hat H$ are obtained by standard numerical diagonalization of \eqref{eq:Hamiltonian} in a truncated finite-dimensional Hilbert space. The truncation is realized by finding the number of states needed to ensure that the lowest-energy eigenvalues and the corresponding eigenvectors, which are involved in the dynamical processes investigated here, are not significantly affected by the truncation. Thereafter, the density matrix in the basis of the system eigenstates is truncated such that all higher-energy eigenstates which are not populated during the dynamical evolution are omitted.

%%%%%%%%%%%%%%%%%%%%%%%%

\subsection{Energy levels and effective interaction strengths}
\label{sec:EnergyLevels}

\begin{figure*}
\centering
\includegraphics[width = \linewidth]{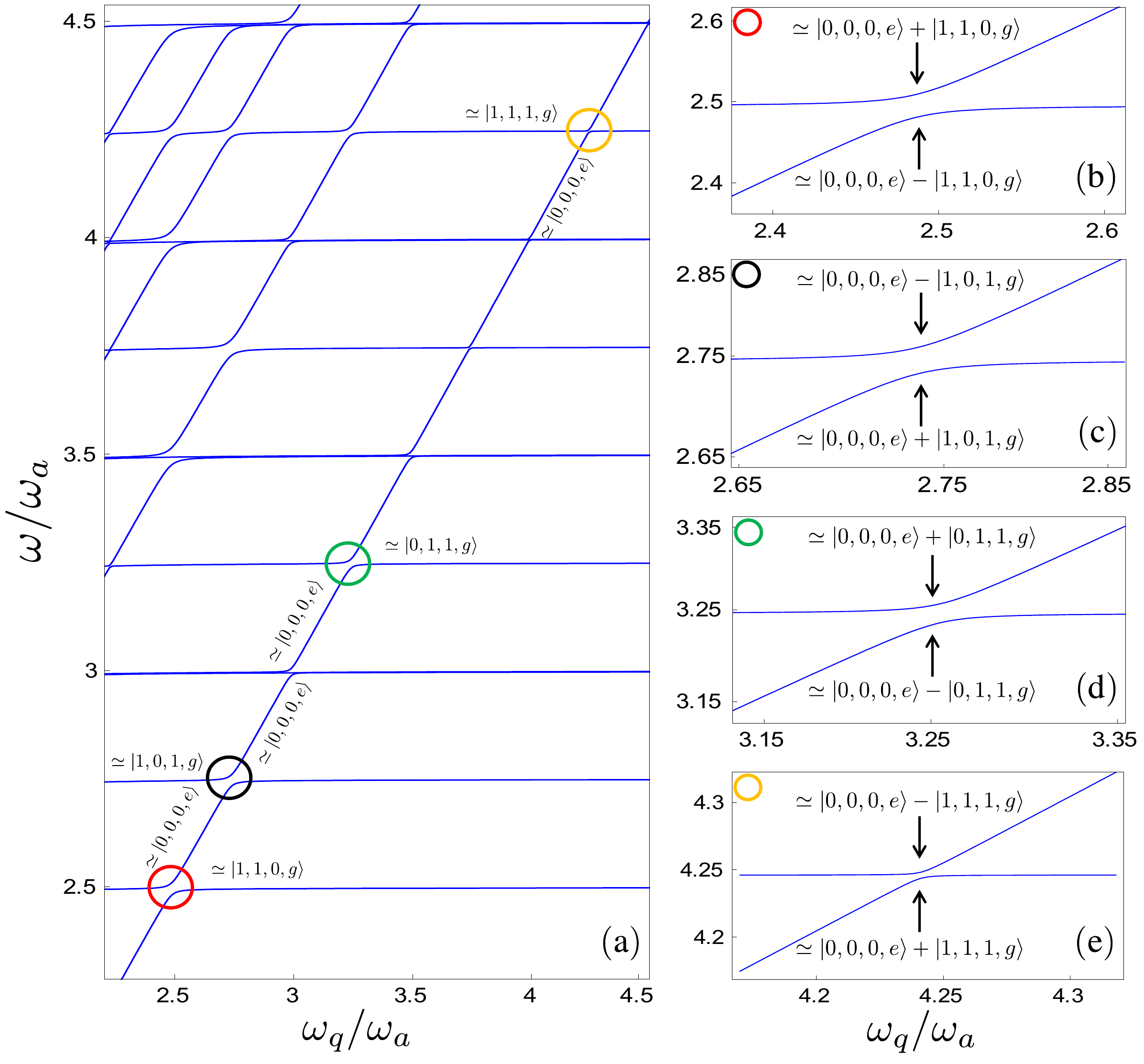}
\caption{Energy levels and avoided level crossings for our system.
(a) The relevant energy levels for our entanglement-generation protocol, normalized by $\omega_a$ and plotted as a function of the qubit frequency $\omega_q$. The transition frequencies of the three resonator modes are kept fixed. The plot is obtained by numerical diagonalization of the Hamiltonian in \eqref{eq:Hamiltonian}. All parameters used for the calculation are provided in the text. Panels
(b), (c), (d), and (e) are zoom-ins of the areas marked by red, black, green, and yellow circles, respectively, in panel (a). These avoided level crossings, which occur when the qubit frequency equals the sum of some of the resonator-mode frequencies, indicate an effective coherent coupling between qubit and photonic states, which we can use for entanglement generation. 
\label{fig:EnergyLevelsAntiCrossings}}
\end{figure*}

To show the basic mechanism for our proposed entanglement-generation scheme, we plot in \figref{fig:EnergyLevelsAntiCrossings}(a) some of the lowest energy levels of our system as a function of the qubit frequency $\omega_q$. The parameters used in the plot are $\omega_{\rm b} = 1.5 \omega_{\rm a}$, $\omega_{\rm c} = 1.75 \omega_{\rm a}$, $g = 0.1 \omega_{\rm a}$, and $\theta = \pi / 6$.

%===================================================
%===================================================

The lowest-energy horizontal line in the plot corresponds to the state with one photon each in the first two resonator modes, no photon in the third resonator mode, and the qubit in its ground state. We denote this state by
\be
\ket{\psi_1, g} = \ket{1, 1, 0, g},
\ee
where on the right-hand side the first three entries are the number of photons in the resonator modes $a$, $b$, and $c$, respectively, and the last entry is the qubit state. To distinguish the qubit state from the photonic states, we hereafter denote the qubit ground state $\ket{g}$ and the excited state $\ket{e}$. Adopting this notation, the second, fourth and eighth horizontal lines correspond to, respectively, 
\bea
\ket{\psi_2, g} &=& \ket{1, 0, 1, g}, \\
\ket{\psi_3, g} &=& \ket{0, 1, 1, g}, \\
\ket{\psi_4, g} &=& \ket{1, 1, 1, g}. 
\eea

We note that these eigenstates can differ from the bare eigenstates due to the dressing effects induced by the counter-rotating terms in the interaction part of the Hamiltonian in \eqref{eq:Hamiltonian}. These differences between bare and physical states occur, more or less, for all the energy eigenstates~\cite{DiStefano2017}. For example, the bare state $\ket{\psi_1} \ket{g}$, describing the excitation of the first two resonator modes in the absence of interaction with the qubit, differs from the dressed state $\ket{\psi_1, g}$ corresponding to the excitation of the first two \textit{physical} resonator modes in the presence of interaction with the qubit. A signature of this dressing is the slight difference between the sum of the bare frequencies $\omega_a + \omega_b = 2.5 \omega_a$ and the lowest horizontal energy level ($\omega \approx 2.5 \omega_a$) displayed in \figref{fig:EnergyLevelsAntiCrossings}(a).

As $\omega_{\rm q}$ increases in \figref{fig:EnergyLevelsAntiCrossings}(a), the energy level associated with the state
\be
\ket{\psi_0, e} = \ket{0, 0, 0, e}
\ee
rises to meet the energy levels corresponding to, from left to right, $\ket{\psi_1, g}$, $\ket{\psi_2, g}$, $\ket{\psi_3, g}$ and $\ket{\psi_4, g}$. The result is several avoided level crossings, marked by colored circles in the plot.

%%%%%%%%%%%%

\subsubsection{Second-order processes: Bell states}

In Figs.~\ref{fig:EnergyLevelsAntiCrossings}(b)-(d), we show enlarged views of the regions marked by the red, black, and green circles in \figref{fig:EnergyLevelsAntiCrossings}(a). These avoided level crossings arise due to coherent coupling between the state $\ket{\psi_0, e}$ and the states $\ket{\psi_1, g}$, $\ket{\psi_2, g}$, and $\ket{\psi_3, g}$, respectively, and occur at the points where the qubit frequency equals the sum of two of the resonator-mode frequencies. Explicitly, we have the resonance conditions 
\be
\omega_q \simeq \omega_a + \omega_b
\ee
in \figref{fig:EnergyLevelsAntiCrossings}(b),
\be
\omega_q \simeq \omega_a + \omega_c
\ee
in \figref{fig:EnergyLevelsAntiCrossings}(c), and
\be
\omega_{\rm q} \simeq \omega_{\rm b} + \omega_{\rm c}
\ee
in \figref{fig:EnergyLevelsAntiCrossings}(d).

The essential point for entanglement generation is that the coherent coupling makes it such that, when the level splitting of the avoided level crossing is at its minimum, the eigenstates of the system are symmetric and antisymmetric superpositions of the states $\ket{\psi_0, e}$ and $\ket{\psi_n, g}$ ($n = 1, 2, 3$). This is confirmed by numerical calculations. Thus, if we first initialize the system in $\ket{\psi_0, e}$, and then tune the qubit frequency such that $\ket{\psi_0, e}$ becomes resonant with $\ket{\psi_n, g}$ ($n = 1, 2, 3$), we will observe Rabi oscillations back and forth between $\ket{\psi_0, e}$ and $\ket{\psi_n, g}$. By initializing the system in a superposition of $\ket{\psi_0, e}$ and $\ket{0, 0, 0, g}$, this allows us to create Bell states for photons in two resonator modes, as explained further below in \secref{sec:Protocol}.

The coherent coupling at the avoided level crossings is due to a second-order process involving both the longitudinal and transversal coupling terms in \eqref{eq:Hamiltonian}. A detailed illustration of this second-order process can be found in \appref{app:CalculationsBell}. The minimum level splitting at the avoided level crossing is determined by $g_{\rm eff}^{\rm (B)}$, the strength of the effective coupling between the states $\ket{\psi_0, e}$ and $\ket{\psi_n, g}$ induced by the second-order process. This coupling strength also sets the timescale for the Rabi oscillations between these states, and thus also the timescale for the entanglement generation in our protocol.

A good approximation of the effective coupling strength can be calculated analytically using second-order perturbation theory, considering all possible paths between the initial state $\ket{i} \equiv \ket{\psi_0} \ket{e}$ and the final state $\ket{f} \equiv \ket{\psi_n} \ket{g}$ ($n = 1, 2, 3$), or vice versa. The details of these calculations are presented in \appref{app:CalculationsBell}. For $n = 1$, the result is 
\bea
g_{\rm eff}^{\rm (B)} = - \frac{g_a g_b \mleft( \omega_a + \omega_b \mright) \sin 2 \theta}{\omega_a \omega_b}.
\label{eq:GEffB}
\eea
This result illustrates why USC is required for our protocol to work well. Since the effective coupling is due to a second-order process, it scales as $g^2/\omega$, and would thus become prohibitively small if $g \ll \omega$.

\begin{figure}
\centering
\includegraphics[width = \linewidth]{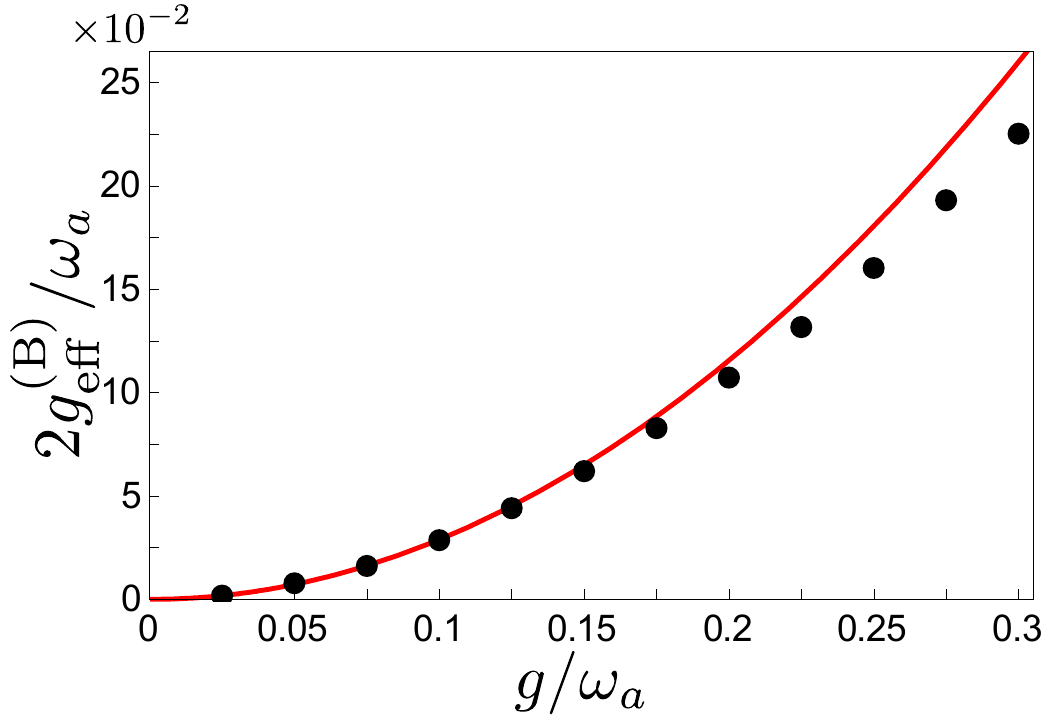}
\caption{A comparison between the numerically calculated normalized effective coupling strength $g_{\rm eff}^{\rm (B)} / \omega_{\rm a}$ (black points) and the corresponding analytical result [\eqref{eq:GEffB}] from second-order perturbation theory (red curve).  We used $g_{\rm a} = g_{\rm b} = g, g_{\rm c} = 0$; all other parameters are the same as in \figref{fig:EnergyLevelsAntiCrossings}.
\label{fig:GEffBNumericsAnalytics}}
\end{figure}

Since the perturbation theory uses $g/\omega$ as its small parameter, and we want to increase $g/\omega$ to generate entanglement faster, we check numerically how well this analytical result for the effective coupling strength agrees with the true value. The result is plotted in \figref{fig:GEffBNumericsAnalytics}. The plot shows $g_{\rm eff}^{(B)}$ as a function of $g / \omega_a$ for $g_a = g_b = g$ and $g_c = 0$. The red curve is the analytical result from \eqref{eq:GEffB} and the black dots are the results from the numerical diagonalization of the Hamiltonian in \eqref{eq:Hamiltonian}. We see that the analytical result remains a very good approximation for normalized interaction strengths $g / \omega_{\rm a} \lesssim 0.2$.

%===================================================
%===================================================

%%%%%%%%%%%%

\subsubsection{Third-order process: GHZ state}

Panel (e) in \figref{fig:EnergyLevelsAntiCrossings} shows an enlarged view of the region marked by a yellow circle in \figref{fig:EnergyLevelsAntiCrossings}(a). This level crossing arises due to a third-order process creating a coherent coupling between the states $\ket{\psi_0, e}$ and $\ket{\psi_4, g}$, and occurs at the point where $\omega_q \simeq \omega_a + \omega_b + \omega_c$. As before, numerical calculations confirm that the eigenstates of the system at the point where the level splitting is at its minimum are the symmetric and antisymmetric superpositions of the states $\ket{\psi_0, e}$ and $\ket{\psi_4, g}$. This means that we can initialize the system in a superposition of $\ket{\psi_0, e}$ and $\ket{0, 0, 0, g}$, and then transfer the qubit excitation to the three resonator modes to create a GHZ state for the photons in these modes, as explained in more detail below in \secref{sec:Protocol}.

In this case, the level splitting is smaller than before, since the process is of a higher order than before. The third-order process responsible for the effective coupling does not require longitudinal coupling, since $\ket{\psi_0, e}$ and $\ket{\psi_4, g}$ have the same parity. A detailed illustration showing all transition paths that contribute to the effective coupling $g_{\rm eff}^{\rm (G)}$ is given in \appref{app:CalculationsGHZ}. From third-order perturbation theory with $\theta = 0$ in \eqref{eq:Hamiltonian}, we find
\be
g_{\rm eff}^{\rm (G)} = - \frac{4 g_a g_b g_c \mleft( \omega_a + \omega_b + \omega_c \mright)}{\mleft( \omega_a + \omega_b \mright) \mleft( \omega_a + \omega_c \mright) \mleft( \omega_b + \omega_c \mright)}.
\label{eq:GEffG}
\ee
The details of the calculations leading to this result are presented in \appref{app:CalculationsGHZ}. As expected for a third-order process, the effective coupling scales as $g^3/\omega^2$, showing that being in the USC regime is more important for generation of GHZ states than for generation of Bell states.

\begin{figure}
\centering
\includegraphics[width = \linewidth]{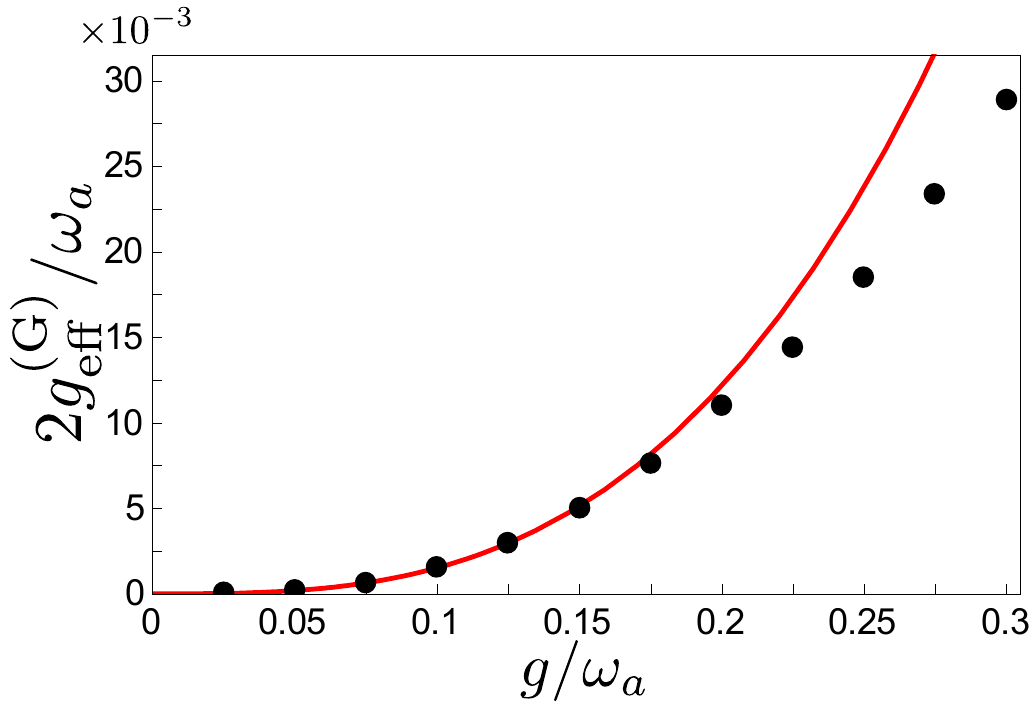}
\caption{A comparison between the numerically calculated normalized effective coupling strength $g_{\rm eff}^{\rm (G)} / \omega_{\rm a}$ (black points) and the corresponding analytical result [\eqref{eq:GEffG}] from third-order perturbation theory (red curve). We used $g_{\rm a} = g_{\rm b} =  g_{\rm c} = g$ and $\theta = 0$; all other parameters are the same as in \figref{fig:EnergyLevelsAntiCrossings}.
\label{fig:GEffGNumericsAnalytics}}
\end{figure}

Just as for $g_{\rm eff}^{\rm (B)}$ above, we compare the analytical result in \eqref{eq:GEffG} with full numerical calculations to see for which parameters the perturbation theory gives a good approximation of the true value for $g_{\rm eff}^{\rm (G)}$. The result of this comparison is plotted in \figref{fig:GEffGNumericsAnalytics}. Just as in \figref{fig:GEffBNumericsAnalytics}, the red curve is the analytical result and the black dots are the results from numerical diagonalization of the Hamiltonian. We note that the agreement between the perturbation-theory result and the correct value remains very good when $g / \omega_{\rm a} \lesssim 0.2$.

%===================================================
%===================================================

%%%%%%%%%%%%%%%%%%%%%%%%%%%%%%%%%%%%%%%%%%%%%%%%%

\section{Results}
\label{sec:Results}

In this section, we first present the details of our entanglement-generation protocol. We then present results from numerical simulations of the protocol for both Bell and GHZ states, using experimentally feasible parameters and exploring the effect of losses on the fidelity of the protocol.

%%%%%%%%%%%%%%%%%%%%%%%

%===================================================
\begin{table*}
\centering
\caption{The three Bell states whose generation we simulate in \figref{fig:BellTimeEvolutionFidelity}, the resonance conditions for their creation, and the qubit-photon coupling strengths used in the simulations. \label{tab:BellResonanceConditions}}
\renewcommand{\arraystretch}{1.7}
\renewcommand{\tabcolsep}{0.2cm}
\begin{tabular}{|c|c|c|c|}
\hline
\textbf{Bell State} & \textbf{Resonance condition} & \textbf{Qubit-photon couplings} \\
\hline
$B_{110} = \frac{1}{\sqrt{2}} \mleft (\ket{\psi_0} + e^{i \phi} \ket{\psi_1} \mright) \ket{g}$ & $\omega_q \simeq \omega_a + \omega_b$ & $g_a = g_b = g, \:\: g_c = 0$ \\
\hline
$B_{101} = \frac{1}{\sqrt{2}} \mleft( \ket{\psi_0} + e^{i \phi} \ket{\psi_2} \mright) \ket{g}$ & $\omega_q \simeq \omega_a + \omega_c$ & $g_a = g_c = g, \:\: g_b = 0$ \\
\hline
$B_{011} = \frac{1}{\sqrt{2}} \mleft( \ket{\psi_0} + e^{i \phi} \ket{\psi_3} \mright) \ket{g}$ & $\omega_q \simeq \omega_b + \omega_c$ & $g_c = g_b = g, \:\: g_a = 0$ \\
\hline
\end{tabular}
\end{table*}
%===================================================

\subsection{Entanglement-generation protocol}
\label{sec:Protocol}

The entanglement-generation protocol essentially consists of two steps. The steps in the protocol are the same for both Bell and GHZ states. The only difference between the two cases is which resonance condition is used; that determines whether photons from two or three resonator modes become entangled.

\textit{Step 1}: We begin with the system in its ground state with the qubit frequency far detuned from any resonance with the resonator-mode frequencies (and their sums), i.e., in the state
\be
\ket{\psi_0}\ket{g} = \ket{0, 0, 0} \ket{g}.
\ee
Then, we rotate the qubit state to a superposition state,
\be
\frac{1}{\sqrt{2}} \ket{\psi_0} \mleft( \ket{g} + \ket{e} \mright).
\ee
The idea of the protocol is to transfer this qubit superposition to several photons.

\textit{Step 2}: We then tune the qubit frequency into resonance with the sum of the frequencies of the resonator modes that we wish to entangle. This will change the system state to
\be
\ket{\psi(t=0)} = \frac{1}{\sqrt{2}} \mleft( \ket{\psi_0, g} + \ket{\psi_0, e} \mright).
\label{eq:Step2T0}
\ee
Let us assume that we have tuned the qubit frequency into resonance with the photonic state $\ket{\psi_n}$ ($n = 1, 2, 3, 4$), i.e., to one of the marked avoided level crossings in \figref{fig:EnergyLevelsAntiCrossings}. As we showed in \secref{sec:EnergyLevels}, this will create an effective coupling of strength $g_{\rm eff}^{\rm (B/G)}$ between the states $\ket{\psi_0, e}$ and $\ket{\psi_n, g}$. The state $\ket{\psi_0, g}$ is not affected. The system state will thus undergo Rabi oscillations and evolve in time as
\bea
\ket{\psi(t)} &=& \frac{1}{\sqrt{2}} \Big[ \ket{\psi_0, g} + \cos \mleft( g_{\rm eff}^{\rm (B/G)} t \mright) \ket{\psi_0, e} \nn\\
&&- i \sin \mleft( g_{\rm eff}^{\rm (B/G)} t \mright) \ket{\psi_n, g} \Big].
\eea
After a time $t = \pi / 2 g_{\rm eff}^{\rm (B/G)}$, we detune the qubit far from the resonance. This leaves the system in the state
\be
\frac{1}{\sqrt{2}} \mleft( \ket{\psi_0} + e^{i\phi} \ket{\psi_n} \mright) \ket{g},
\ee
where $\phi \in \mathbb{R}$. This is an entangled state of photons in the resonator modes. The qubit is no longer entangled with the photons. For $n = 1, 2, 3$, the entangled photonic state is a Bell state for the photons in resonator modes $a$ and $b$, $a$ and $c$, or $b$ and $c$, respectively. For $n = 4$, the entangled photonic state is a GHZ state involving all three resonator modes. We note again that the case $n = 4$ does not require the longitudinal coupling term in \eqref{eq:Hamiltonian}, since this process conserves the parity of the number of excitations in the system.

%%%%%%%%%%%%%%%%%%%%%%%

\subsection{Numerical simulations for the Bell states}
\label{sec:ResultsBell}

\begin{figure*}
\centering
\includegraphics[width = \linewidth]{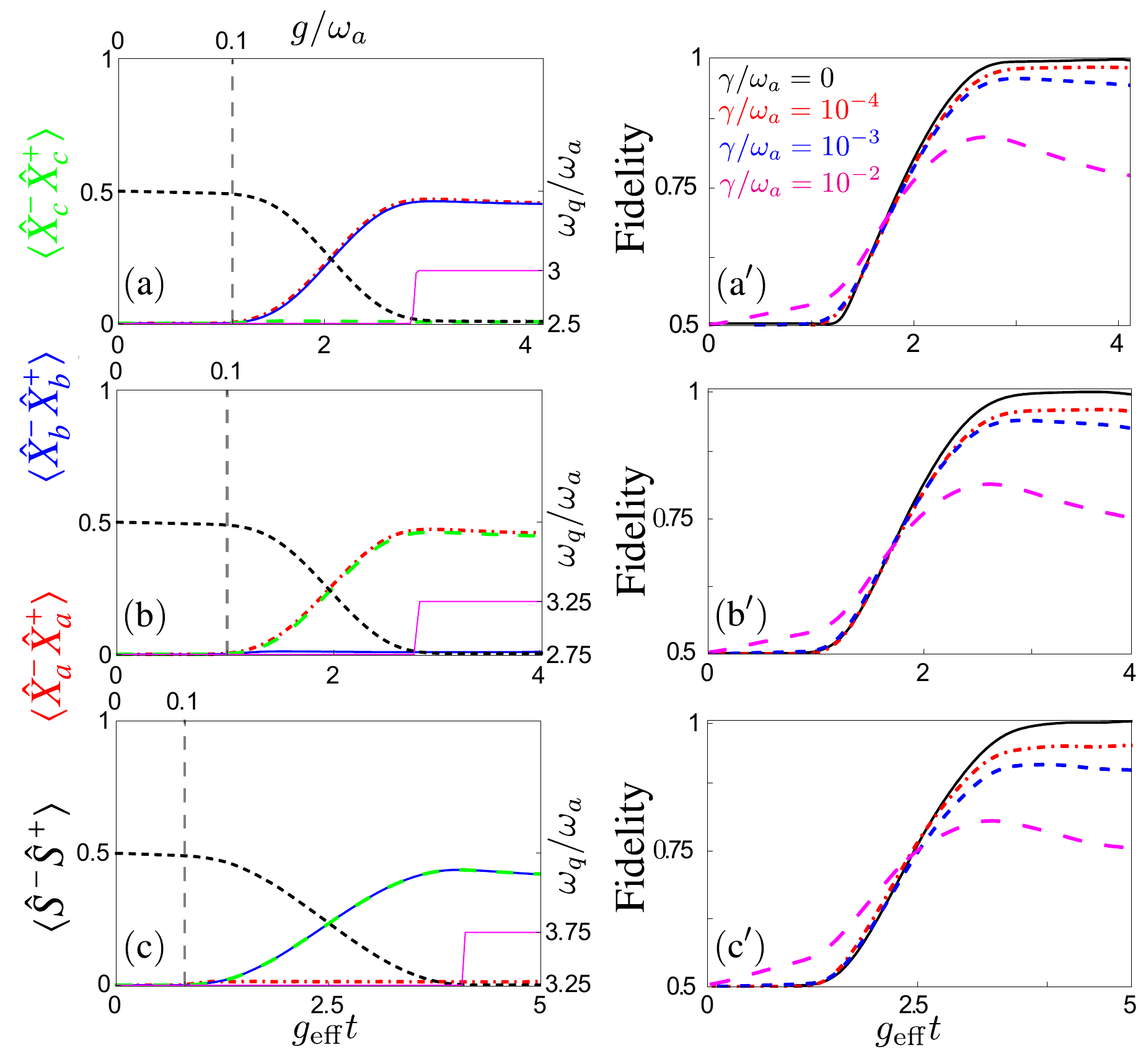}
\caption{Simulations of the protocol for Bell-state generation under the influence of decoherence.
The left panels show the number of excitations in resonator mode $a$, $\expec{\hat X_a^- \hat X_a^+}$ (red dashed-dotted curves), in resonator mode $b$, $\expec{\hat X_b^- \hat X_b^+}$ (blue solid curves), in resonator mode $c$, $\expec{\hat X_c^- \hat X_c^+}$ (green large-dashed curve), and in the qubit, $\expec{\hat S^- \hat S^+}$ (black dashed curve). These are plotted as a function of time during the creation of the Bell states (a) $B_{110}$, (b) $B_{101}$, and (c) $B_{011}$ (see \tabref{tab:BellResonanceConditions} for definition of this notation). In panels (a)-(c), the vertical grey dashed lines indicate when the coupling between the qubit (which starts in a superposition state) and the relevant resonator modes is turned on. The pink solid curve shows how the qubit frequency $\omega_q$ is tuned during the protocol. This is given by the smoothed step function $\omega_q (t) = \omega_{q, i} + \delta \omega_q \mleft\{ \sin^2 \mleft[ A (t - t_i) \mright] \Theta (t - t_i) + \sin^2 \mleft[ A (t - t_f) \mright] \Theta (t - t_f) \mright\} $, where $\omega_{q, i}$ is the initial qubit frequency, $\delta \omega_q$ is the change of the qubit frequency, $\Theta$ is the Heaviside step function, $t_i$ is the time when the qubit frequency starts to change, $t_f = t_i + \pi / (2 A)$, and $A$ is a frequency setting the smoothness.
Panels (a')-(c') are plots of the fidelities for producing the desired entangled states from panels (a)-(c) given different decoherence rates. All parameters for the simulations are given in the text.
\label{fig:BellTimeEvolutionFidelity}}
\end{figure*}

Here we simulate our protocol for Bell-state generation, taking into account losses. We do this by solving the master equation in \eqref{eq:ME}. The results are presented in \figref{fig:BellTimeEvolutionFidelity}. We use the same parameters for the resonator-mode and qubit frequencies and coupling strength as in \figref{fig:EnergyLevelsAntiCrossings}. To simplify the simulations, we start with the qubit already in a superposition state and tuned to the desired resonance, but with all couplings turned off. At the time marked by the vertical grey dashed line in panels (a)-(c) in \figref{fig:BellTimeEvolutionFidelity}, we turn on the coupling to the two resonator modes that we wish to create a Bell state for, i.e., at this point, we start from the state in \eqref{eq:Step2T0}. After a time $t \approx \pi / 2 g_{\rm eff}^{\rm (B)}$, we detune the qubit frequency from the resonance [pink curve in panels (a)-(c) in \figref{fig:BellTimeEvolutionFidelity}]. For clarity, we list in \tabref{tab:BellResonanceConditions} the three Bell states together with the corresponding resonance conditions and couplings turned on in the simulations.

%===================================================
%===================================================

In panels (a)-(c) in \figref{fig:BellTimeEvolutionFidelity}, we show the time evolution for the qubit and resonator-mode populations during the whole protocol. Note that the populations must be calculated using the dressed operators given in \eqref{eq:DressedOperators}. Panel (a) is for the generation of the Bell state with photons in modes $a$ and $b$, panel (b) is the result for photons in modes $a$ and $c$, and panel (c) is for the case with photons in modes $b$ and $c$. In all these panels, we use the decoherence parameters $\gamma = 10^{-3} \omega_{\rm a}$ and $\kappa_{\rm a} = \kappa_{\rm b} = \kappa_{\rm c}= \gamma / 2$.

To test the robustness of our Bell-state-generation protocol, we repeated these simulations for different values of the decoherence rate $\gamma$ and calculated the fidelity $\mathcal{F} = \sqrt{\brakket{\psi}{\rho(t)}{\psi}}$ for the desired entangled state $\ket{\psi}$. The results are shown in panels (a')-(c') in \figref{fig:BellTimeEvolutionFidelity}. We observe that the protocol works well as long as $\gamma / \omega_a \lesssim 10^{-3}$, producing entangled states with a fidelity of 90\% or above. For larger decoherence rates, the fidelity becomes markedly lower, since we then enter a regime where $\gamma$ starts to become comparable to $g_{\rm eff}^{\rm (B)}$.

%%%%%%%%%%%%%%%%%%%%%%%

\subsection{Numerical simulations for the GHZ state}
\label{sec:ResultsGHZ}

\begin{figure}
\centering
\includegraphics[width = \linewidth]{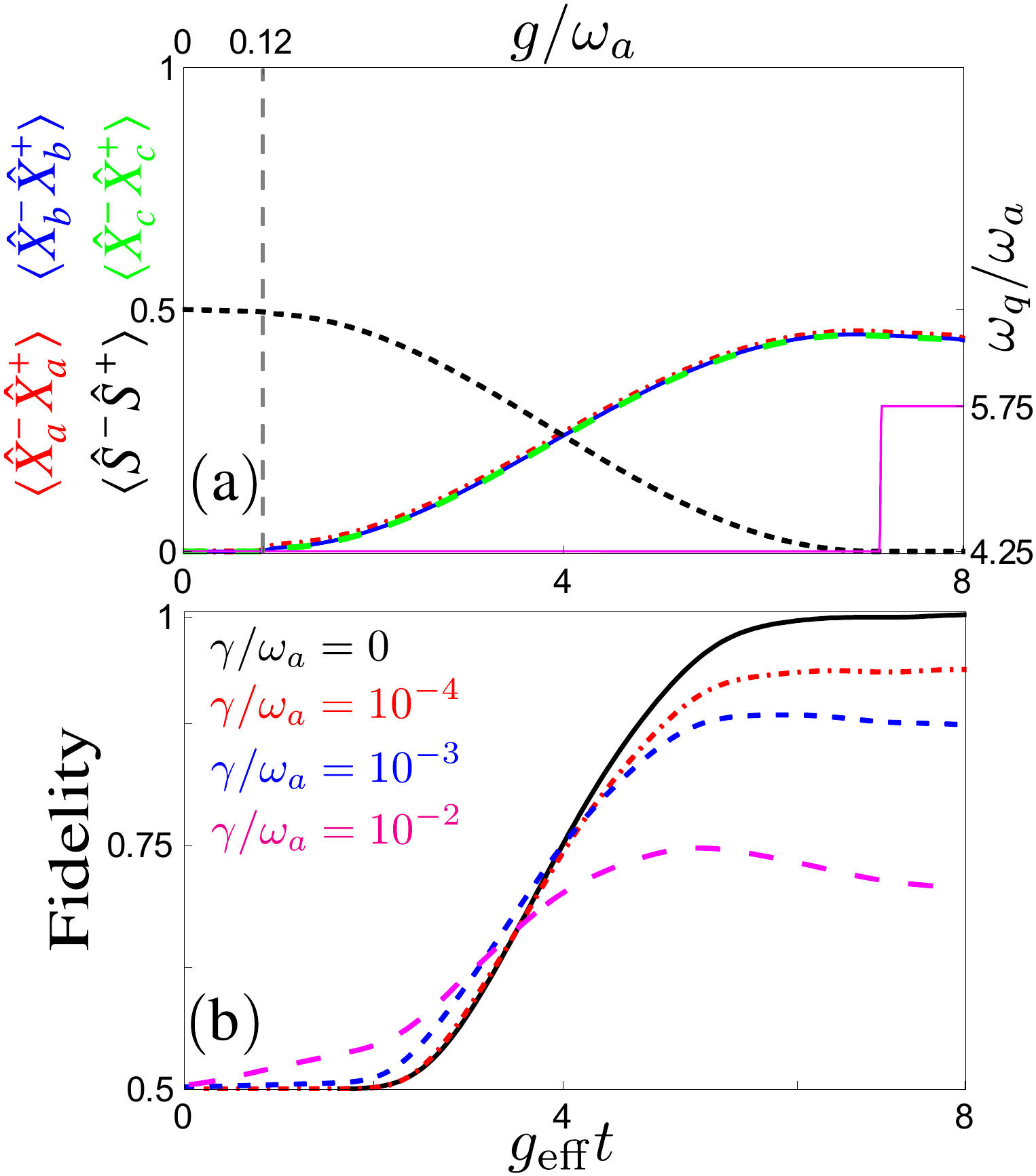}
\caption{Simulations of the protocol for GHZ-state generation under the influence of decoherence.
(a) The number of excitations in resonator mode $a$, $\expec{\hat X_a^- \hat X_a^+}$ (red dashed-dotted curves), in resonator mode $b$, $\expec{\hat X_b^- \hat X_b^+}$ (blue solid curves), in resonator mode $c$, $\expec{\hat X_c^- \hat X_c^+}$ (green large-dashed curve), and in the qubit, $\expec{\hat S^- \hat S^+}$ (black dashed curve). These are plotted as a function of time during the creation of the GHZ state. The vertical grey dashed line indicates when the coupling between the qubit (which starts in a superposition state) and the resonator modes is turned on. The pink solid curve shows how the qubit frequency $\omega_q$ is tuned during the protocol; the form of this curve is the same as in \figref{fig:BellTimeEvolutionFidelity}.
(b) Fidelities for producing the desired GHZ state from panel (a) given different decoherence rates. All parameters for the simulations are given in the text. \label{fig:TimeEvolutionFidelityGHZ}}
\end{figure}

We now perform numerical simulations also for the GHZ-state-generation protocol with losses included. We do this in the same way as for the Bell states in \secref{sec:ResultsBell}. The results are plotted in \figref{fig:TimeEvolutionFidelityGHZ}. The resonator-mode frequencies are the same as in the previous plots, while the qubit frequency is tuned to the resonance for the GHZ state, $\omega_{\rm q} \simeq \omega_{\rm a} + \omega_{\rm b} + \omega_{\rm c}$. Since a longitudinal coupling is not required to generate the GHZ state, we set $\theta = 0$. We use a slightly higher coupling strength, $g = 0.12 \omega_{\rm a}$, since $g_{\rm eff}^{\rm (G)}$ otherwise would be much weaker than $g_{\rm eff}^{\rm (B)}$. When the coupling is turned on, marked by the vertical dashed grey line in panel (a), we set $g_a = g_b = g_c = g$. Decoherence is included with qubit and resonator losses at the same levels as in \figref{fig:BellTimeEvolutionFidelity}.

%===================================================
%===================================================

In panel (a) in \figref{fig:TimeEvolutionFidelityGHZ}, we show the time evolution for the qubit and resonator-mode populations during the whole protocol. In panel (b), we plot the fidelity for the GHZ state as a function of time for different decoherence rates. Again, the protocol works well when $\gamma / \omega_a \lesssim 10^{-3}$; for larger decoherence rates, the fidelity declines since $\gamma$ starts to become comparable to $g_{\rm eff}^{\rm (G)}$.

%%%%%%%%%%%%%%%%%%%%%%%

\subsection{Experimental feasibility}
\label{sec:ExpFeasibility}

Here we briefly comment on the experimental feasibility of our entanglement-generation protocol. As described in the introduction, circuit-QED systems are the only experimental setups so far to demonstrate USC with single atoms. We therefore limit the discussion here to parameters in circuit-QED experiments, although our protocol may become possible to implement in other systems in the future.

As we saw in Secs.~\ref{sec:ResultsBell} and \ref{sec:ResultsGHZ}, the essential requirement for generating entangled photonic states with high fidelity using our protocol is that the effective coupling $g_{\rm eff}^{\rm (B/G)}$ is clearly larger than the decoherence rates in the system. Since circuit-QED systems can reach the USC regime $g / \omega > 0.1$, they can surely realize $g_{\rm eff}^{\rm (B/G)} / \omega \sim 10^{-3} - 10^{-2}$ [compare Eqs.~(\ref{eq:GEffB}) and (\ref{eq:GEffG})]. When it comes to decoherence rates, flux qubits have demonstrated relaxation rates as low as $\gamma / \omega \sim 10^{-6}$~\cite{Stern2014, Orgiazzi2016, Yan2016}. Similarly, superconducting transmission-line resonators with relaxation rates on the order of $\kappa / \omega \sim 10^{-6}$~\cite{Megrant2012} have been fabricated; three-dimensional resonators for circuit QED can even reach $\kappa / \omega \sim 10^{-8}$~\cite{Reagor2013}. We therefore expect that our scheme can create photonic Bell and GHZ states with high fidelity using existing technology.

%%%%%%%%%%%%%%%%%%%%%%%%%%%%%%%%%%%%%%%%%%%%%%%%%

\section{Conclusion and outlook}
\label{sec:ConclusionOutlook}

We have presented a simple protocol for the deterministic preparation of photonic Bell and GHZ states. We do this in a setup consisting of three resonator modes, all of different frequencies, coupled to a single qubit. The protocol relies on this coupling between light and matter being ultrastrong, which enables higher-order processes that do not conserve the number of excitations in the system. Using this property, a superposition state prepared for the qubit can be transferred to multiple photons in the resonator modes. Our protocol is versatile, since the same setup, and the same steps, can be used to generate all these maximally entangled states. The only thing that changes depending on which entangled state we wish to create, is which frequency we tune the qubit to. The resonance condition we need is that the qubit frequency equals the sum of the frequencies of the resonator modes that are to be entangled.

We have shown that our protocol is ready to be implemented in circuit-QED experiments using state-of-the-art technology. Seeing as USC is being achieved in more and more experimental systems, we expect that our protocol will be useful in other systems as well in the future. We believe that our protocol could be useful as a method to distribute entanglement at a node in a quantum network.

It is clear that our protocol can be extended to create entangled states involving four or more photons of different frequencies. This simply requires adding more resonator modes and tuning the qubit frequency to equal the sum of these resonator-mode frequencies. However, the effective coupling strength, which determines how fast we can transfer the superposition state from the qubit to the photons, will become weaker the more resonator modes are involved. This is because populating more photonic modes requires a process of higher order, and the effective coupling scales as $\mleft( g / \omega \mright)^{n-1}$ for an $n$th-order process~\cite{Kockum2017a}. Such a weak effective coupling will only give an entangled state with low fidelity unless the decoherence rates in the system are even lower. 

It is also interesting to think about whether our protocol can be extended to create other types of entangled states. For three photons, the GHZ and W states are the only two distinct types of maximally entangled states~\cite{Dur2000}. However, it is not clear how the idea of our protocol could be used to create a W state for photons of different frequencies, since this requires creating a superposition of three different states, all with different energies.

%%%%%%%%%%%%%%%%%%%%%%%%%%%%%%%%%%%%%%%%%%%%%%%%%

\acknowledgments

We thank Salvatore Savasta for useful discussions. F.N.~acknowledges support from the MURI Center for Dynamic Magneto-Optics via the Air Force Office of Scientific Research (AFOSR) award No.~FA9550-14-1-0040, the Army Research Office (ARO) under grant No.~W911NF-18-1-0358, the Asian Office of Aerospace Research and Development (AOARD) grant No.~FA2386-18-1-4045, the Japan Science and Technology Agency (JST) through the ImPACT program and CREST Grant No.~JPMJCR1676, the Japan Society for the Promotion of Science (JSPS) through the JSPS-RFBR grant No.~17-52-50023 and the JSPS-FWO grant No.~VS.059.18N, the RIKEN-AIST Challenge Research Fund, and the John Templeton Foundation. A.F.K.~acknowledges partial support from a JSPS Postdoctoral Fellowship for Overseas Researchers (P15750).

%%%%%%%%%%%%%%%%%%%%%%%%%%%%%%%%%%%%%%%%%%%%%%%%%

\appendix

%%%%%%%%%%%%%%%%%%%%%%%%%%%%%%%%%%%%%%%%%%%%%%%%%

\section{Analytical calculation of the effective coupling for generating Bell states}
\label{app:CalculationsBell}

In this appendix, we calculate the effective coupling rates $g_{\rm eff}^{\rm (B)}$ for the processes that we use to generate entangled Bell states between photons in two resonator modes.

%%%%%%%%%%%%%%%%%%%%%%%

\subsection{System}

For the Bell-state-generation, we only need to consider two resonator modes, with transition frequencies $\omega_a$ and $\omega_b$, respectively, both ultrastrongly coupled to a qubit with transition frequency $\omega_q$. The relevant system Hamiltonian is obtained from \eqref{eq:Hamiltonian}, neglecting the third resonator mode:
\bea
\hat H &=& \omega_a \hat a^\dag \hat a + \omega_b \hat b^\dag \hat b + \frac{\omega_q}{2} \sz \nn\\
&&+ \mleft[ g_a \mleft( \hat a^\dag + \hat a \mright) + g_b \mleft( \hat b^\dag + \hat b \mright) \mright]  \mleft( \sx \cos \theta + \sz \sin \theta \mright) , \nn\\
\label{eq:HBell}
\eea
where $\hat a$ ($\hat a^\dag$) is the annihilation (creation) operator of the first resonator mode, $\hat b$ ($\hat b^\dag$)
is the annihilation (creation) operator of the second resonator mode, $\sz$ and $\sx$ are the Pauli matrices for the qubit, $g_a$ ($g_b$) is the coupling between the first (second) resonator and the qubit, and $\theta$ is an angle parameterizing the amount of longitudinal and transversal coupling. The longitudinal coupling term in this generalized quantum Rabi Hamiltonian is necessary for the process we have in mind, since that process neither conserves the number of excitations in the system nor their parity.

%%%%%%%%%%%%%%%%%%%%%%%

\subsection{Transition paths and perturbation theory}

\begin{figure}
\centering
\includegraphics[width = 0.8 \linewidth]{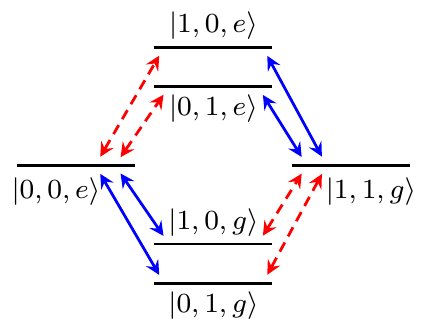}
\caption{The four second-order paths connecting the states $\ket{0, 0, e}$ and $\ket{1, 1, g}$. Transitions that do not conserve the number of excitations in the system are marked by dashed lines while transitions that conserve the number of excitations are marked by solid lines. Red lines mark transitions mediated by the $\sz$ part of the coupling and blue lines mark transitions mediated by the $\sx$ part of the coupling. To set the energy levels, we have used the parameter values $\omega_q = 2.5 \omega_b$ and $\omega_a = 1.5 \omega_b$.
\label{fig:TransitionPaths00e11g}}
\end{figure}

To generate the Bell state, we need a transition between the states $\ket{0, 0, e}$ and $\ket{1, 1, g}$. These two states are connected via a second-order process. There are four paths connecting the states to this order, as shown in \figref{fig:TransitionPaths00e11g}.

%===================================================
%===================================================

We can treat the interaction part of the Hamiltonian in \eqref{eq:HBell},
\be
\hat V = \mleft [ g_{\rm a} \mleft( \hat a^\dag + \hat a \mright) + g_{\rm b} \mleft( \hat b^\dag + \hat b \mright) \mright] \mleft( \sx \cos \theta + \sz \sin \theta \mright),
\ee
as a perturbation, provided that $g_a, g_b \ll \omega_a, \omega_b, \omega_q$. Writing
\be
V_{nm} = \brakket{n}{\hat V}{m},
\ee
from second-order perturbation theory we have that the effective coupling between the initial state $\ket{i} = \ket{0, 0, e}$ and the final state $\ket{f} = \ket{1, 1, g}$ is given by
\be
g_{\rm eff} = \sum_{n} \frac{V_{fn} V_{ni}}{\mleft( E_i - E_n \mright)},
\label{eq:GeffPerturbationFormula}
\ee
where the sum goes over all paths shown in \figref{fig:TransitionPaths00e11g}. Note that $E_i = \frac{1}{2} \omega_q = E_f = \omega_a + \omega_b - \frac{1}{2} \omega _q$ when we are at the resonance $\omega_q = \omega_a + \omega_b$, which is the case we consider here. The contributions from the four transition paths add up to
\bea
g_{\rm eff}^{\rm (B)} &=& - g_a g_b \sin \theta \cos \theta \mleft( \frac{1}{\omega_a} + \frac{1}{\omega_b} + \frac{1}{\omega_q - \omega_a} + \frac{1}{\omega_q - \omega_b} \mright) \nn\\
&=& - g_a g_b \sin 2 \theta \mleft( \frac{1}{\omega_a} + \frac{1}{\omega_b} \mright) \nn\\
&=& - \frac{g_a g_b \mleft( \omega_a + \omega_b \mright) \sin 2 \theta}{\omega_a \omega_b}.
\label{eq:Geff2Modes1}
\eea

Looking at the denominator of \eqref{eq:Geff2Modes1}, we see that $g_{\rm eff} \to \infty$ when $\omega_a \to 0$ or $\omega_a \to 0$, i.e., when the qubit becomes resonant with one of resonator modes. The perturbation theory is not valid around those points, nor is it valid when $\omega_a \approx \omega_b$, since the states $\ket{2, 0, g}$ and $\ket{0, 2, g}$ then would have approximately the same energy as the initial and final states in the process we considered here. For similar reasons, we also want to avoid that $\omega_q = n \omega_a$, $\omega_q = m \omega_b$, and $\omega_a = k \omega_b$, where $n, m, k$ are integers. Finally, we note that since the effective coupling is proportional to $\sin 2 \theta$, it will reach its maximum value when $\theta = \pi/4 + n \pi/2$.

%%%%%%%%%%%%%%%%%%%%%%%%%%%%%%%%%%%%%%%%%%%%%%%%%

\section{Analytical calculation of the effective coupling for generating GHZ states}
\label{app:CalculationsGHZ}

In this appendix, we calculate the effective coupling rates $g_{\rm eff}^{\rm (G)}$ for the process that we use to generate an entangled GHZ state between photons in three resonator modes.

%%%%%%%%%%%%%%%%%%%%%%%

\subsection{System}

Our system now consists of three resonator modes, with transition frequencies $\omega_a$, $\omega_b$, and $\omega_c$, respectively, all ultrastrongly coupled to a qubit with transition frequency $\omega_q$. The relevant system Hamiltonian is obtained from \eqref{eq:Hamiltonian} with $\theta = 0$:
\bea
\hat H &=& \omega_{\rm a} \hat a^\dag \hat a + \omega_{\rm b} \hat b^\dag \hat b + \omega_{\rm c} \hat c^\dag \hat c + \frac{\omega_{\rm q}}{2} \sz \nn\\
&&+ \sx \mleft [ g_{\rm a} \mleft( \hat a^\dag + \hat a \mright) + g_{\rm b} \mleft( \hat b^\dag + \hat b \mright) + g_{\rm c} \mleft( \hat c^\dag + \hat c \mright) \mright] , \quad
\label{eq:HGHZ}
\eea
where $\hat a$ ($\hat a^\dag$) is the annihilation (creation) operator of the first resonator mode, $\hat b$ ($\hat b^\dag$) is the annihilation (creation) operator of the second resonator mode, $\hat c$ ($\hat c^\dag$) is the annihilation (creation) operator of the third resonator mode, $\sz$ and $\sx$ are the Pauli matrices for the qubit, and $g_j$ is the coupling between resonator $j$ and the qubit. For this process, the standard quantum Rabi Hamiltonian with only transversal coupling ($\theta = 0$), expanded to include multiple resonator modes, is sufficient, since the process needed for the GHZ-state generation conserves the parity of the number of excitations.

%%%%%%%%%%%%%%%%%%%%%%%

\subsection{Transition paths and perturbation theory}

\begin{figure}
\centering
\includegraphics[width = \linewidth]{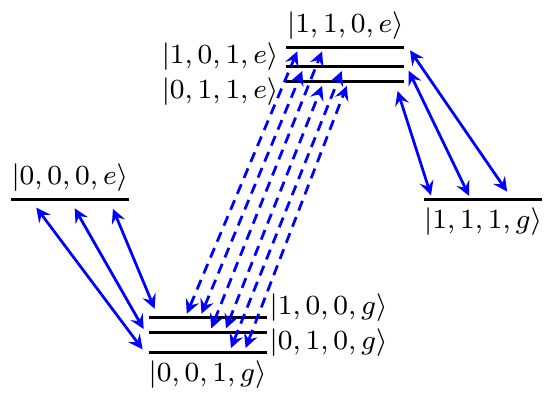}
\caption{The six third-order paths connecting the states $\ket{0, 0, 0, e}$ and $\ket{1, 1, 1, g}$. Transitions that do not conserve the number of excitations in the system are marked by dashed lines while transitions that conserve the number of excitations are marked by solid lines. To set the energy levels, we have used the parameter values $\omega_q = 4.1 \omega_c$, $\omega_a = 1.7 \omega_c$, and $\omega_b = 1.4 \omega_c$.
\label{fig:TransitionPaths000e111g}}
\end{figure}

To generate the GHZ state, we need a transition between the states $\ket{0, 0, 0, e}$ and $\ket{1, 1, 1, g}$. These two states are connected via a third-order process. There are six paths connecting the states to this order, as shown in \figref{fig:TransitionPaths000e111g}.

%===================================================
%===================================================

We can treat the interaction part of the Hamiltonian in \eqref{eq:HGHZ},
\be
\hat V = \sx \mleft[ g_{\rm a} \mleft( \hat a^\dag + \hat a \mright) + g_{\rm b} \mleft( \hat b^\dag + \hat b \mright) + g_{\rm c} \mleft( \hat c^\dag + \hat c \mright) \mright] ,
\ee
as a perturbation, provided that $g_a, g_b, g_c \ll \omega_a, \omega_b, \omega_c, \omega_q$. From third-order perturbation theory, we have that the effective coupling between the initial state $\ket{i} = \ket{0, 0, 0, e}$ and the final state $\ket{f} = \ket{1, 1, 1, g}$ is given by
\be
g_{\rm eff} = \sum_{n,m} \frac{V_{fn} V_{nm} V_{mi}}{\mleft( E_i - E_n \mright) \mleft( E_i - E_m \mright)},
\ee
where the sum goes over all paths shown in \figref{fig:TransitionPaths000e111g}. Note that $E_i = \frac{1}{2} \omega_q = E_f = \omega_a + \omega_b + \omega_c - \frac{1}{2} \omega _q$ when we are at the resonance $\omega_q = \omega_a + \omega_b + \omega_c$, which is the case we consider here. The contributions from the six transition paths add up to

\begin{widetext}

\bea
g_{\rm eff}^{(G)} &=& - g_a g_b g_c \Bigg[ \frac{1}{\mleft( \omega_a + \omega_b \mright) \mleft( \omega_q - \omega_a \mright)} + \frac{1}{\mleft( \omega_a + \omega_b \mright) \mleft( \omega_q - \omega_b \mright)} + \frac{1}{\mleft( \omega_a + \omega_c \mright) \mleft( \omega_q - \omega_a \mright)} \nn\\
&&+ \frac{1}{\mleft( \omega_a + \omega_c \mright) \mleft( \omega_q - \omega_c \mright)} + \frac{1}{\mleft( \omega_b + \omega_c \mright) \mleft( \omega_q - \omega_b \mright)} + \frac{1}{\mleft( \omega_b + \omega_c \mright) \mleft( \omega_q - \omega_c \mright)} \Bigg] \nn\\
&=& - 2 g_a g_b g_c \mleft[ \frac{1}{\mleft( \omega_a + \omega_b \mright) \mleft( \omega_b + \omega_c \mright)} + \frac{1}{\mleft( \omega_a + \omega_c \mright) \mleft( \omega_b + \omega_c \mright)} + \frac{1}{\mleft( \omega_a + \omega_b \mright) \mleft( \omega_a + \omega_c \mright)} \mright] \nn\\
&=& - \frac{4 g_a g_b g_c \mleft( \omega_a + \omega_b + \omega_c \mright)}{\mleft( \omega_a + \omega_b \mright) \mleft( \omega_a + \omega_c \mright) \mleft( \omega_b + \omega_c \mright)}.
\label{eq:Geff3Modes}
\eea

\end{widetext}

Looking at the denominator of \eqref{eq:Geff3Modes}, we see that $g_{\rm eff} \to \infty$ if two of the three resonator frequencies go to zero, \textit{i.e.}, when the qubit becomes resonant with one of resonator modes. The perturbation theory is not valid around those points. Nor is it valid when additional states, containing two or more photons in one of the resonator modes, have approximately the same energy as the initial and final states in the process we considered here.

%========== Bibliography =============

\bibliography{ReferencesBellGHZ}

\end{document}